\documentstyle[aps,preprint]{revtex}
\begin{document}
\title{Quasiparticle photoemission intensity in doped two-dimensional quantum
antiferromagnets}
\author{F. Lema and A. A. Aligia}
\address{Centro At\'{o}mico Bariloche and Instituto Balseiro,\\
Comisi\'on Nacional de Energ\'{\i}a At\'omica,\\
8400 Bariloche, Argentina}
\maketitle

\begin{abstract}
Using the self-consistent Born approximation, and the corresponding wave
function of the magnetic polaron, we calculate the quasiparticle weight
corresponding to destruction of a real electron (in contrast to creation of
a spinless holon), as a funtion of wave vector for one hole in a generalized 
$t-J$ model and the strong coupling limit of a generalized Hubbard model.
The results are in excellent agreement with those obtained by exact
diagonalization of a sufficiently large cluster. Only the Hubbard weigth
compares very well with photoemission measurements in Sr$_2$CuO$_2$Cl$_2$.
\end{abstract}


\newpage

The problem of a single hole in an antiferromagnetic background has been a
subject of considerable interest since the discovery of high-$T_c$ systems .
One of the most powerful tools for this study is the self-consistent Born
approximation (SCBA) \cite{sch,kan,mar,liu}. Excellent agreement has been
obtained between the position of the lowest pole of the holon Green function
of the SCBA and the quasiparticle dispersion obtained by exact
diagonalization of small systems \cite{mar,liu,xia}. An important advance in
the understanding of the SCBA has been the explicit construction of the
corresponding wave function by Reiter \cite{rei}.

The interest on the problem has been revived by recent angle-resolved
photoemission experiments on insulating Sr$_2$CuO$_2$Cl$_2$, in which the
hole dispersion and quasiparticle weight have been measured \cite{wel}.
While it was clear that the ``bare'' $t-J$ model was unable to explain the
observed dispersion, several works have appeared fitting the experimental
dispersion using generalized $t-J$ models \cite{xia,naz,bel}, a generalized
Hubbard model \cite{lem} and the spin-fermion (or Kondo-Heisenberg) model
for the cuprates \cite{sta}. Except for the fact that the band width is $%
\sim 10\%$ narrower than the experimental result if the experimental value
of $J$ is taken \cite{ero}, the generalized $t-J$ model including hopping to
second and third NN and the three-site term $t^{\prime \prime }$, reproduces
well the experimental dispersion \cite{xia,bel} and also other properties of
the spin-fermion and three-band Hubbard models \cite{ero}. A consistent
picture of the observed spin and charge excitations has been obtained using
a generalized one-band Hubbard model \cite{lem}.

However, very little attention has been devoted to the explanation of the
intensity of the observed quasiparticle peaks. This task is difficult for
the following reasons: i) exact results for quasiparticle intensities in
sufficiently large clusters (containing more than 16 unit cells, as
discussed below) exist only for the ``bare '' $t-J$ model and only at a few
wave vectors. ii) The SCBA provides the Green function of the spinless
holon, while the Green function of the real particles contain spin-wave
excitations and simple decoupling approximations do not provide reasonable
results. The holon weights are the same for wave vectors differing in ($\pi
,\pi )$ contrary to experiment. iii) While a lot of work has been devoted to
the mapping of the three-band Hubbard {\em model} for the cuprates to
low-energy effective {\em models, }less attention has been devoted to the
mapping of the corresponding {\em \ operators} \cite{ero,bat,fei}. This
information as well as the photoionization cross sections for Cu and O are
necessary if accurate weights are wished.

In this paper we calculate the photoemission quasiparticle weight for
removing an electron, as a function of wave vector in generalized $t-J$ and
strong-coupling Hubbard models, using the SCBA {\em and} the wave function
of the polaron \cite{rei}. The Hamiltonian has the form 
\begin{eqnarray}
H &=&-\sum_{i\delta \sigma }t_\delta c_{i+\delta \sigma }^{\dag }c_{i\sigma
}-t^{\prime \prime }\sum_{i\eta \neq \eta ^{\prime }\sigma }c_{i+\eta
^{\prime }\sigma }^{\dag }c_{i\eta \sigma }({\frac 12}-2{\bf S}_i\cdot {\bf S%
}_{i+\eta })  \nonumber \\
&&+{\frac J2}\sum_{i\eta \sigma }({\bf S}_i\cdot {\bf S}_{i+\eta }-{\frac 14}%
n_in_{i+\eta }).  \label{f1}
\end{eqnarray}

The first term contains hopping to first, second and third nearest neighbors
(NN) with parameters $t_1,~t_2,~t_3$ respectively. The first NN of site $i$
are labeled as $i+\eta $. Eq. (\ref{f1}) is obtained from a standard
canonical transformation of a Hubbard model with hoppings $t_1,~t_2,~t_3$,
if (complicated) terms smaller than $t^{\prime \prime }=t^2/U$ are neglected 
\cite{bal}. The difference between generalized $t-J$ and strong-coupling
Hubbard models is the meaning of the operator $c_{i\sigma }$, as explained
below. The Hamiltonian can be written in terms of spinless fermions and
spin-wave operators \cite{sch,kan,mar,liu,bal}. We adopt the procedure and
notation used by Mart\'{\i}nez and Horsch \cite{mar}, slightly generalized
to include second and third NN hoppings and the three-site term \cite{bal}:
The sublattice A is defined as that of positive magnetization. The spins of
sublattice B are rotated 180$^{\circ }$ around the x axis. In this way the
Neel state is converted into a fully polarized ferromagnetic state,
restoring the translational symmetry of the nonmagnetic state at the price
of losing the conservation of spin. Then, the $c_{i\uparrow }^{}$ operator
is defined as a spinless holon creation operator $h_i^{\dagger }$, while $%
c_{i\downarrow }$ becomes a composite operator involving a local spin
deviation $a_i$. The result of both operations is the following
representation:

\begin{eqnarray}
c_{i\uparrow } &=&h_i^{\dagger },\,\,\,\,\,\,\,\,\,c_{i\downarrow
}=h_i^{\dagger }a_i\,,\,\,\,\,\,\text{if }\,\,i\in \,\text{A}  \nonumber \\
c_{i\uparrow } &=&h_i^{\dagger }a_i,\,\,\,c_{i\downarrow }=h_i^{\dagger
}\,\,,\,\,\,\,\,\,\,\,\,\,\text{if}\,\,\,\,i\in \,\text{B.}  \label{f2}
\end{eqnarray}
In the exchange part (last term of Eq. (\ref{f1})) the fermion occupation
numbers are averaged and the bosonic quadratic part is diagonalized by a
standard canonical transformation:

\begin{equation}
\alpha _q=u_qa_q-v_qa_{-q}^{\dagger },  \label{f3}
\end{equation}
where $u_q^{2_{}}=v_q^2+1=1/2+1/(2\nu _q)$ $\,\,,\,\,\,\,\,\nu _q=(1-\gamma
_q^2)^{1/2}\,\,,$ $u_q>0,$ sgn($v_q$) = sgn($\gamma _q$), and $\gamma _q=$
(cos $q_x$ + cos $q_y$)/2. Retaining only linear terms in spin deviations
for the rest of Eq. (\ref{f1}), the Hamiltonian becomes:

\begin{eqnarray}
H &=&E_J^0+\sum_q\omega _q\alpha _q^{\dagger }\alpha _q+\sum_k\epsilon
_kh_k^{\dagger }h_k  \nonumber \\
&&+\frac{4t_1}{\sqrt{N}}\sum_{kq}M(k,q)(h_k^{\dagger }h_{k-q}\alpha _q+\text{%
H.c.}),  \label{f4}
\end{eqnarray}
where $E_J^0$ is a constant, $\omega _q=2J\,\nu _q,$ $\epsilon
_k=(t_2+2(1-x)t^{\prime \prime })\epsilon _2(k)+(t_3+(1-x)t^{\prime \prime
})\epsilon _1(2k)$ and $M(k,q)=(u_q\gamma _{k-q}+v_q\gamma _k),$ with $%
\epsilon _1(k)=4\gamma _k$ and $\epsilon _2(k)=4$ cos $k_x$ cos $k_y$. In
the present case, the doping $x=0$. The constraint that at the same site
there cannot be both a hole and a spin deviation is neglected since it
does not affect the results for motion of a hole in a {\em quantum} antiferromagnet
\cite{mar}. The holon Green function $G_h(k,\omega )$
is obtained from the self-consistent solution of the following two equations:

\begin{eqnarray}
\Sigma (k,\omega ) &=&\frac{4t_1}N\sum_qM^2(k,q)G_h(k-q,\omega -\omega _q) 
\nonumber \\
\,\,\,\,\,\,\,\,\,\,\,\,\,\,\,\,\,\,\,\,G^{-1}(k,\omega ) &=&\omega
-\epsilon _k-\Sigma (k,\omega )+i\epsilon .  \label{f5}
\end{eqnarray}

We have solved Eqs. (\ref{f5}) in clusters of $16\times 16$ and $20\times 20$
sites. In order to obtain accurate values of the holon quasiparticle weight $%
Z_h$, we have discretized the frequencies in intervals of $\Delta \omega
=10^{-4}t_1$ and have taken the small imaginary part $\epsilon =5\Delta
\omega $. As an alternative method to that used by Liu and Manousakis \cite
{liu}, we have fitted the part of the spectral weight nearest to the
quasiparticle peak by a sum of several Lorentzian functions. The resulting
width of the quasiparticle peak was practically identical to $2\epsilon $
and from its integrated weight we determined $Z_h$. We have verified that
using this method there are practically no finite-size effects in our
clusters.

In the sudden approximation, the angle-resolved photoemission spectrum is
proportional to the spectral density of states for Cu and O at wave vector $k
$. These in turn are related to the imaginary part of the Green function for
the generalized $t-J$ operator $c_{k\sigma }$ or the generalized Hubbard
operator $\tilde{c}_{k\sigma }$ through a low-energy reduction procedure 
\cite{ero,fei}. In linear order in $1/U$, the well known procedure of the
canonical transformation \cite{bat,esk} applied to the generalized Hubbard
model, in the subspace of no double occupancy, leads to:

\begin{equation}
\tilde{c}_{i\sigma }=c_{i\sigma }+\sum_\delta \frac{t_\delta }U(n_{i\bar{%
\sigma}}c_{i+\delta \sigma }-c_{i\bar{\sigma}}^{\dagger }c_ic_{i+\delta \bar{%
\sigma}}).  \label{f6}
\end{equation}
Calling $\mid 0\rangle $ ($\mid \psi _k\rangle $) the ground state of Eq. (%
\ref{f4}) for the undoped (hole doped with wave vector $k$ ) system, and
using the Lehmann representation of the wave function, one realizes that
while the holon quasiparticle weight is:

\begin{equation}
Z_h(k)=\mid \langle \psi _k\mid h_k^{\dagger }\mid 0\rangle \mid ^2,
\label{f7}
\end{equation}
the weight for emitting a Hubbard electron is:

\begin{equation}
Z_{c\sigma }^{GH}(k)=\mid \langle \psi _k\mid \tilde{c}_{k\sigma }\mid
0\rangle \mid ^2+\mid \langle \psi _{k+Q}\mid \tilde{c}_{k\sigma }\mid
0\rangle \mid ^2,  \label{f8}
\end{equation}
where $Q=(\pi ,\pi )$, and $\mid \psi _k\rangle $ and $\mid \psi
_{k+Q}\rangle $ are the {\em degenerate }eigenstates of lowest energy of Eq.
(\ref{f4}) with a finite overlap with $\tilde{c}_{k\sigma }\mid 0\rangle $ .
The corresponding result for the generalized $t-J$ model $Z_{c\sigma
}^{GtJ}(k)$ is obtained taking infinite $U$. Since $Z_{c\uparrow
}(k)=Z_{c\downarrow }(k)$ we restrict to spin up in the following. The
states $\mid \psi _k\rangle $ can be constructed following the procedure
used by Reiter \cite{rei}. The only change in Eqs. 1 to 10 of Ref. \cite{rei}%
, is that the quasiparticle energy $\lambda _k=$ $\lambda _{k+Q}$ is
replaced by $\lambda _k-\epsilon _k$ in Eqs. 3, 6 and 9, and by $\lambda
_k-\epsilon _{k-q}$ in Eq. 4 . Thus, writing explicitely only the terms with
less than two spin-wave excitations we have:

\begin{eqnarray}
\mid &\psi _k\rangle =A_0(k)h_k^{\dagger }\mid 0\rangle +\frac 1{\sqrt{N}}%
\sum_qA_1(k,q)h_{k-q}^{\dagger }\alpha _q^{\dagger }\mid 0\rangle +..., \label{f9}
\end{eqnarray}
where:
\begin{eqnarray}
A_1(k,q)=4t_1M(k,q)G_h(k-q,\lambda _k-\omega _q)A_0(k).  \label{f9p}
\end{eqnarray}
Using Eqs. (\ref{f2}) and (\ref{f6}) and retaining only terms lines in spin
deviations we obtain:

\begin{eqnarray}
\tilde{c}_{i\uparrow } &=&h_i^{\dagger }-\frac{t_1}U(1-x)\sum_\eta h_{i+\eta
}^{\dagger }a_i^{\dagger }\,,\,\,\,\,\,\text{if }\,\,i\in \,\text{A} 
\nonumber \\
\tilde{c}_{i\uparrow } &=&h_i^{\dagger }a_i+\frac{1-x}U[t_1\sum_\eta
h_{i+\eta }^{\dagger }+\sum_{\delta \neq \eta }t_\delta h_{i+\delta
}^{\dagger }(a_{i+\delta }-a_i)],\,\,\,i\in \,\text{B.}\,  \label{f10}
\end{eqnarray}
The most important correction of order $1/U$ is the first term between
brackets in the second Eq. (\ref{f10}) and reflects the fact that in the
ground state of the undoped Hubbard model, there is a finite double
occupancy at sites B and an electron with spin up can be destroyed there,
leaving a hole in one of its NN (this leads to the second term between
brackets in Eqs. (\ref{f11}) and (\ref{f13})).

Expressing Eqs. (\ref{f10}) in Fourier components, and using $\sum_{i\in 
\text{A(B)}}e^{ikR_i}=(\delta _{k,0}+e^{iQR_i}\delta _{k,Q})N/2$, we obtain:

\begin{equation}
\tilde{c}_{k\uparrow }=\frac 12(1+f(k))(h_k^{\dagger }+s_Ah_{k+Q}^{\dagger
})+\frac 1{2\sqrt{N}}\sum_q(h_{k+q}^{\dagger }-s_Ah_{k+q+Q}^{\dagger
})[(1+g(k,q))a_q-f(k)a_q^{\dagger }],  \label{f11}
\end{equation}
where the phase $s_A=e^{iQR_i}$ with $i\in $ A, and

\begin{equation}
f(k)=\frac{t_1}U(1-x)\epsilon _1(k),\,\,\,\,\,g(k,q)=\frac{1-x}U%
[t_2(\epsilon _2(k)-\epsilon _2(k+q))+t_3(\epsilon _1(2k)-\epsilon
_1(2k+2q))].  \label{f12}
\end{equation}
Using Eqs. (\ref{f3}), (\ref{f7}), (\ref{f8}), (\ref{f9}), (\ref{f9p}) and (\ref{f11}) we
obtain the desired result:

\begin{equation}
\frac{Z_{c\sigma }^{GH}(k)}{Z_h(k)}=\frac 12\mid 1+f(k)+\frac{8t_1}N%
\sum_q^{^{\prime }}M(k,q)G_h(k-q,\lambda _k-\omega
_q)[v_q(1+g(k,q))-u_qf(k)]\mid ^2.  \label{f13}
\end{equation}
The sum is restricted to the magnetic Brillouin zone and the term with $q=0$
is excluded (there are no magnons with $q=0$ or $q=Q$ in the $\mid \psi
_k\rangle $ ). The weight $Z_{c\sigma }^{GtJ}$ for the generalized $t-J$
model operator $c_{i\sigma }$ is given by Eq. (\ref{f13}) with the Hubbard
perturbative corrections $f(k)$ (first NN) and $g(k,q)$ (second and third
NN) set to zero.

In Fig. 1 we compare the weight for the $t-J$ model obtained by exact
diagonalization $Z_{ED}^{tJ}(k)$ in a square lattice of 20 sites \cite{poi}
with our results $Z_{c\sigma }^{tJ}(k)$ for the $20\times 20$ cluster at
equivalent wave vectors. The comparison between exact results for square
clusters of 16, 18, 20 and 26 sites suggest that while the $Z_{ED}^{tJ}(k)$
are nearly 20\% larger for the $4\times 4$ cluster, the finite size effects
are of the order of 5\% for larger clusters \cite{poi}. The agreement between
the exact $Z_{ED}^{tJ}(k)$ and SCBA $Z_{c\sigma }^{tJ}(k)$ results is quite
satisfactory. Note that the very small value of $Z^{tJ}(Q)$ is a severe test
to Eq. (\ref{f13}), since it requires a near cancellation of the different
terms. Instead, the ``bare'' SCBA result satisfies $Z_h(k)=Z_h(k+Q)$ and
cannot reproduce the shape of the exact results.

With the confidence gained by the above comparison, we have calculated the
generalized $t-J$ and Hubbard weights for parameters which fit the observed
quasiparticle dispersion $\lambda _k$ in Sr$_2$CuO$_2$Cl$_2$ \cite{wel}.
There are several choices of $t_2,t_3$ and $t^{\prime \prime }$, including
different signs of $t^{\prime \prime }$ which produce nearly identical
results. We took the parameters of Ref. \cite{xia}. The resulting dispersion
and weights are represented in Fig. 2. Compared with the parameters of Fig.
1, the effects of $t_2,t_3$ and $t^{\prime \prime }$ are dramatic. They push
the $\lambda _k$ towards the incoherent part of the spectrum and reduce
considerably the weights for the lowest $\lambda _k$ (in the electron
representation of Fig. 2). As a consequence, we could not detect
quasiparticles near $k=0,$ $Q$ or $(\pi ,0)$ ($Z_h<10^{-4}$ for these $k$).
Therefore, the corresponding $\lambda _k$ are not represented in Fig. 2. The
weights for the generalized $t-J$ and Hubbard models have significant
differences: in contrast to the results for $t_2=t_3=t^{\prime \prime }=0$
(not shown), $Z_{c\sigma }^{GtJ}(k)$ is larger for $k=(\pi /2+\varepsilon
,\pi /2+\varepsilon )$ than for $k=(\pi /2-\varepsilon ,\pi /2-\varepsilon )$
with small $\varepsilon $. Instead, $Z_{c\sigma }^{GH}(k)$, in agreement
with experiment, is larger inside the non interacting Fermi surface. This
effect is more noticeable for smaller values of $U$ ($t_1/U=0.1$ was taken
in Fig. 2) \cite{esk2}.

In summary, using the SCBA and related wave function, we have calculated the
dispersion and quasiparticle weight for removing a real electron in an
undoped antiferromagnet described by a generalized $t-J$ or a generalized
Hubbard model in the strong coupling limit. The weight for the $t-J$ model
agrees very well with available exact results in sufficiently large
clusters. While the generalized Hubbard can explain well both the measured
dispersion {\em and }weight of the quasiparticle in Sr$_2$CuO$_2$Cl$_2$, the
generalized $t-J$ model, without mapping the electron operators, cannot.

One of us (FL) is supported by the Consejo Nacional de Investigaciones
Cient\'{\i}ficas y T\'{e}cnicas (CONICET), Argentina. (AAA) is partially
supported by CONICET.

{\it Note added}: after submission of this manuscript we became aware of
exact diagonalization results of the $t-J$ model in a square cluster of 32
sites with periodic boundary conditions which has 9 non-equivalent wave vectors  
\cite{leu}. (See Fig. 3). The 
dispersion relation $\lambda_k$ agrees very well with the SCBA results except at the points
$k=(0,0)$, $(\pi/4,\pi/4)$ and $(\pi,\pi/2)$, where finite-size effects are obvious
from the fact that  $\lambda_k \neq \lambda_{k+Q}$. Except at $k=(0,0)$ and $(\pi/4,\pi/4)$,
where the position of $\lambda_k$ affects the quasiparticle weights, these weigths
are in excellent agreement with our results  using Eq. (\ref{f13}).

%\newpage

\section*{Figure Captions}

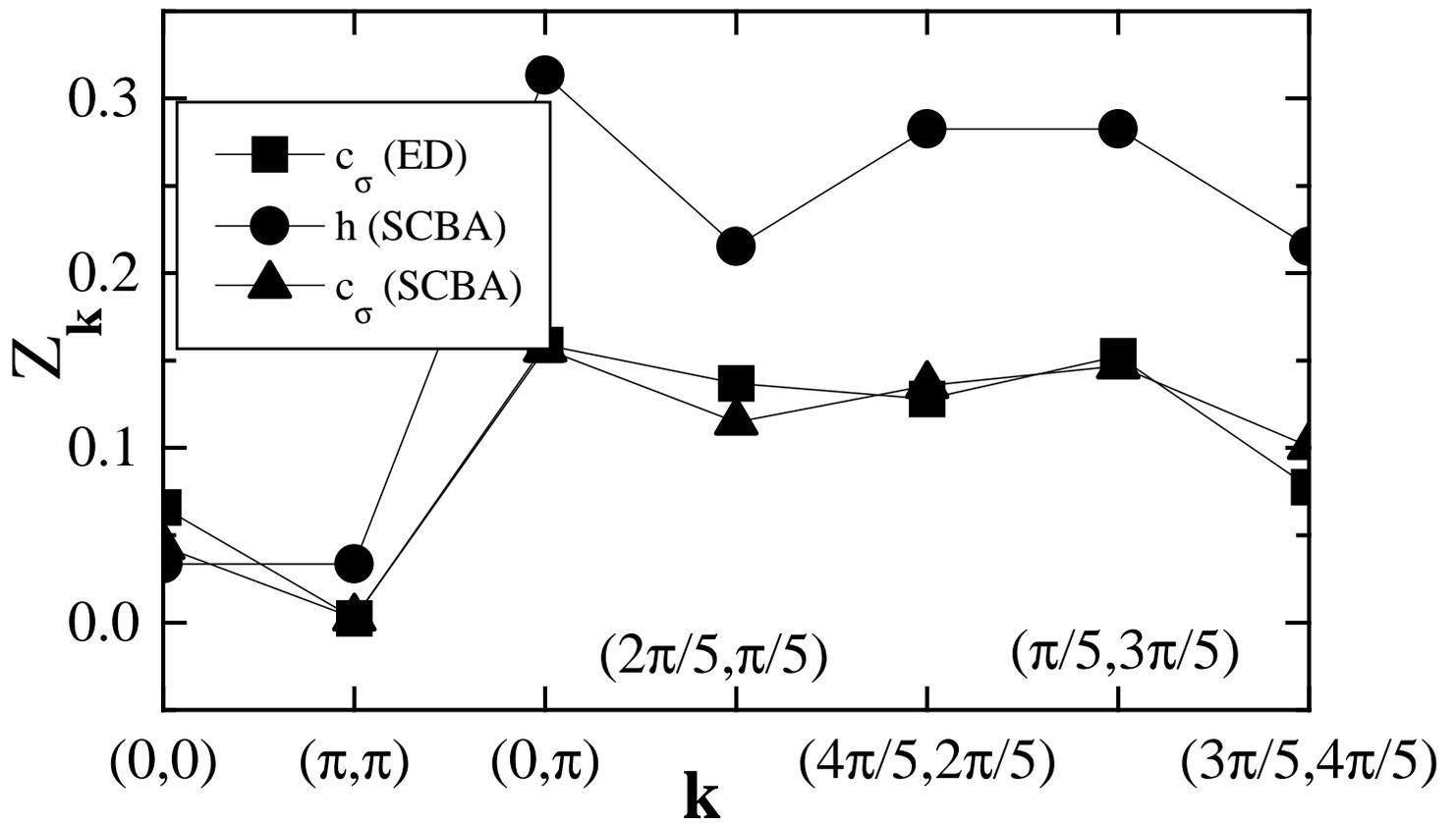
\noindent{\bf Fig.1:} Quasiparticle weight of the $t-J$ model $Z_{c\sigma
}^{tJ}(k)$ calculated with the SCBA in a $20\times 20$ lattice for several
wave vectors (triangles){\bf \ }, compared with exact diagonalization
results in a square cluster of 20 sites $Z_{ED}^{tJ}(k)$ \cite{poi,note}
(squares), and the spinless holon weight $Z_h(k)$ of the SCBA (circles).
Parameters are $t_1=1,~J=0.3,~t_2=t_3=t^{\prime \prime }=0$.

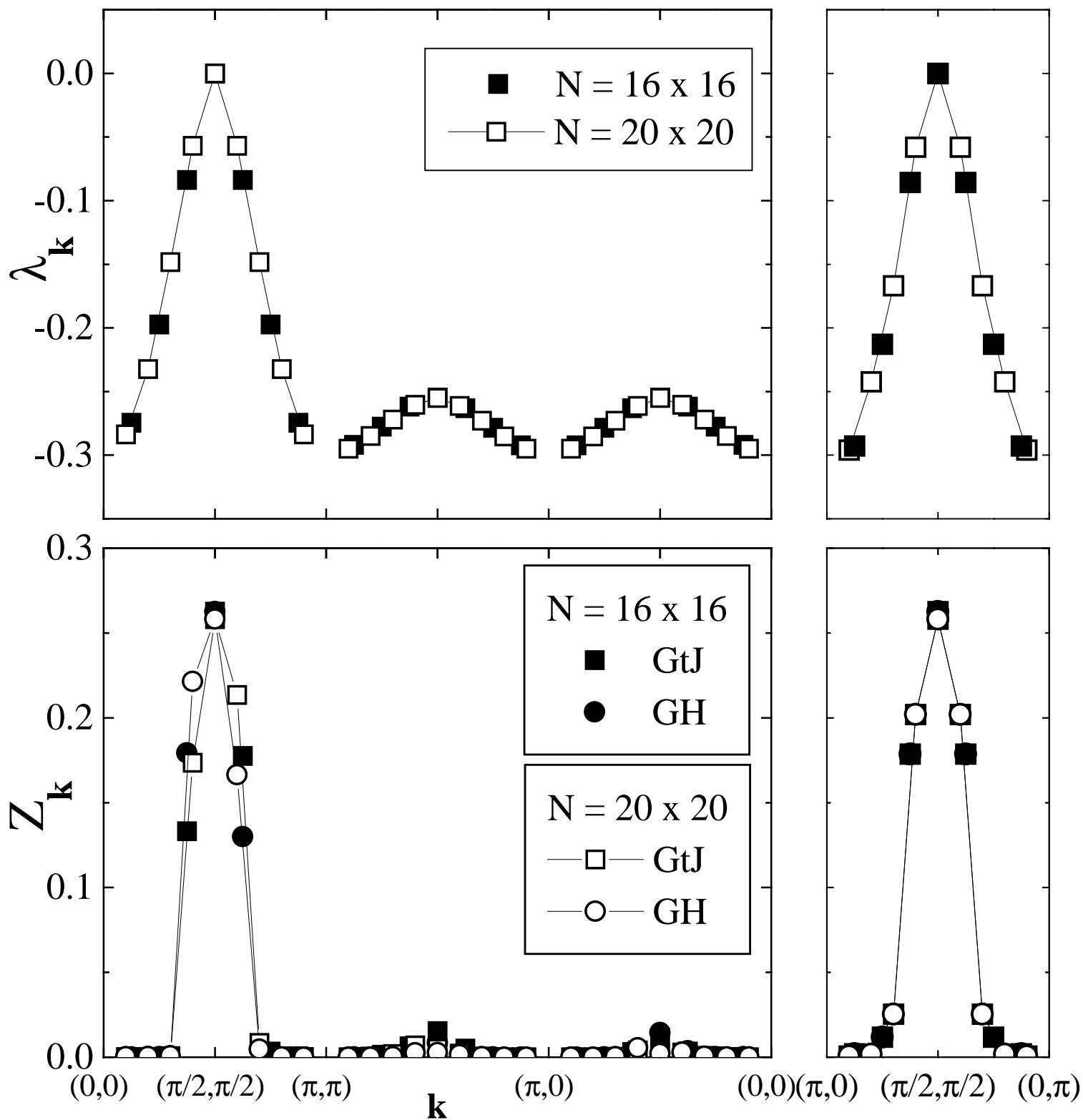
\noindent{\bf Fig.2:} Top: quasiparticle dispersion in clusters of $16\times
16$ (solid symbols) and $20\times 20$ sites (open symbols). Bottom:
corresponding generalized $t-J$ (squares) and generalized Hubbard (circles)
quasiparticle weights. Parameters are: $%
t_1=0.35,~t_2=-0.12,~t_3=0.08,~J=0.15,~t^{\prime \prime }=J/4$ and $U=3.5$  

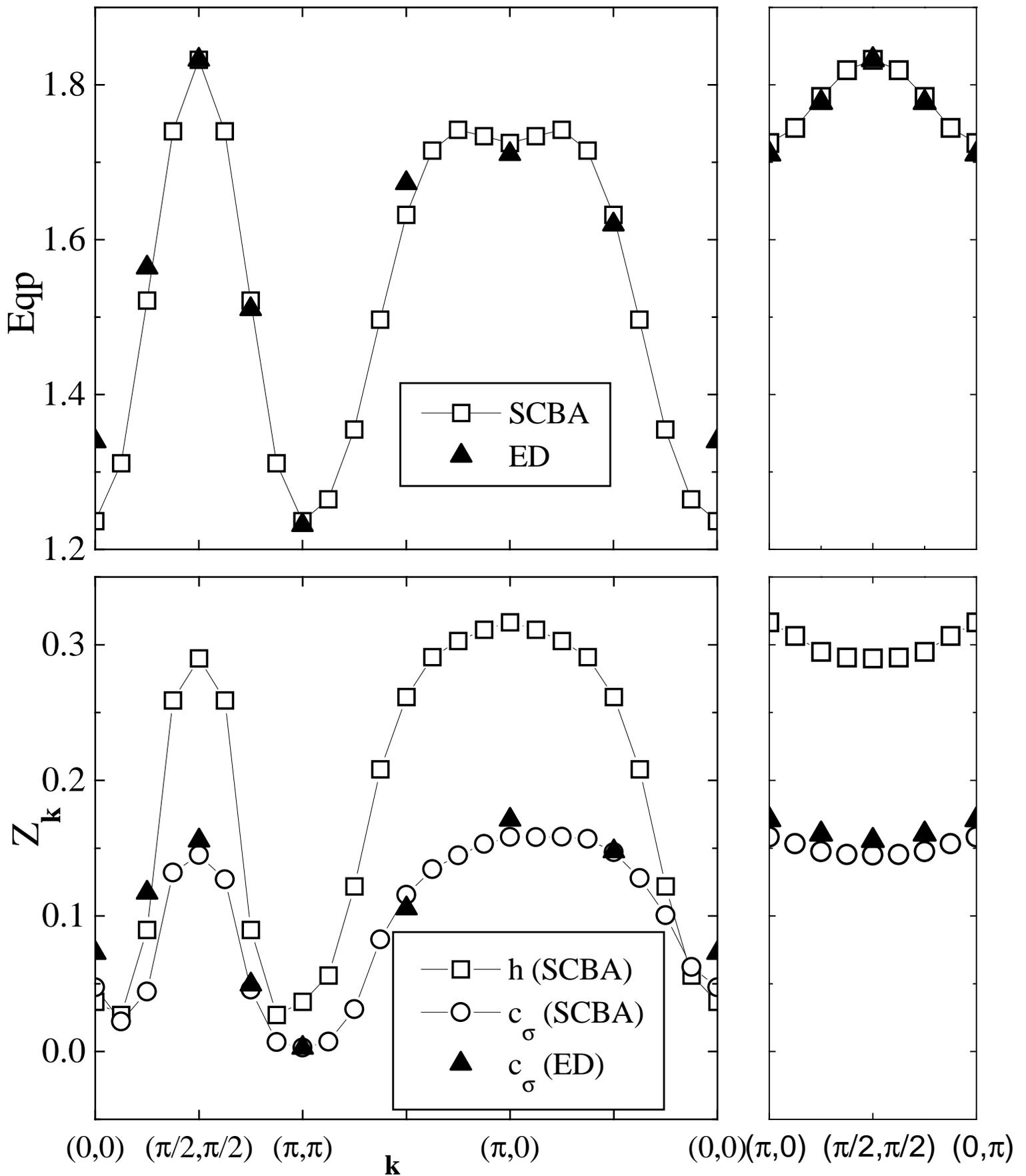
\noindent{\bf Fig.3:} Top: quasiparticle dispersion in the cluster of $16\times
16$ (open squares) compared with the exact diagonalization
results in a square cluster of 32 sites \cite{leu}(solid triangles). Bottom: corresponding 
quasiparticle weights.

\begin{references}
\bibitem{sch}  S. Schmitt-Rink, C.M. Varma, and A.E. Ruckenstein, Phys. Rev.
Lett. {\bf 60,} 2793 (1988).

\bibitem{kan}  C.L. Kane, P.A. Lee, and N. Read, Phys. Rev. B {\bf 39}, 6880
(1988).

\bibitem{mar}  G. Mart\'{\i}nez and P. Horsch, Phys. Rev. B {\bf 44}, 317
(1991).

\bibitem{liu}  Z. Liu and E. Manousakis, Phys. Rev. B {\bf 45}, 2425 (1992).

\bibitem{xia}  T. Xiang and J.M. Wheatley, Phys. Rev. B {\bf 54}, R12653
(1996).

\bibitem{rei}  G.F. Reiter, Phys. Rev. B {\bf 49}, 1536 (1994).

\bibitem{wel}  B.O. Wells, Z.-X. Shen, A. Matsuura, D.M. King, M.A. Kastner,
M. Greven, and R.J. Birgeneau, Phys. Rev. Lett. {\bf 74,} 964 (1995).

\bibitem{naz}  A. Nazarenko, K.J.E. Vos, S. Haas, E. Dagotto, and R.
Gooding, Phys. Rev. B {\bf 51}, 8676 (1995).

\bibitem{bel}  V.I. Belinicher, A.L. Chernyshev, and V.A. Shubin, Phys. Rev.
B {\bf 54}, 14914 (1996).

\bibitem{lem}  F. Lema, J. Eroles, C.D. Batista and E. Gagliano, Phys. Rev.
B (BUR574).

\bibitem{sta}  O.A. Starykh, O.F. de Alcantara Bonfim, and G. Reiter, Phys.
Rev. B {\bf 52}, 12534 (1995).

\bibitem{ero}  J. Eroles, C.D. Batista and A.A. Aligia, Physica C {\bf 261,}
237 (1996); references therein. For large O-O hopping the resulting
three-site term $t^{\prime \prime }$ favors a resonance-valence-bond
superconducting state in the square lattice [C.D. Batista and A.A. Aligia,
Physica C {\bf 261} 237 (1996); J. Low Temp. Phys. {\bf 105}, 591 (1996)]
and shifts towards lower values of $J$ the region of dominant
superconducting correlations in one dimension [F. Lema, C.D. Batista and
A.A. Aligia, Physica C {\bf 259,} 287 (1996)].

\bibitem{poi}  D. Poilblanc, T. Ziman, H.J. Schulz, and E. Dagotto, Phys.
Rev. B {\bf 47}, 14267 (1993).

\bibitem{bat}  C.D. Batista and A.A. Aligia, Phys. Rev. B {\bf 47}, 8929
(1993).

\bibitem{fei}  L. Feiner, Phys. Rev. B {\bf 48}, 16857 (1993).

\bibitem{bal}  J. Bala, A.M. Ole\'{s}, and J. Zaanen, Phys. Rev. B {\bf 52},
4597 (1995).

\bibitem{esk}  H. Eskes, A.M. Ole\'{s}, M.B.J. Meinders, and W. Stephan,
Phys. Rev. B {\bf 50}, 17980 (1994); references therein.        

\bibitem{note}  When comparing our results with those of Poilblanc {\it et
al.}, it should be taken into account that they took the opposite sign of $%
t_1$ (positive for holes), what is equivalent to a shift in ${\bf Q}=(\pi
,\pi )$ of all wave vectors, and they summed over both spins, introducing a
factor 2 with respect to our results. To compare with experiment the sign of 
$t_1$ should be determined by the mapping procedure from the three-band
Hubbard model $H_{3b}$ with original phases\cite{ero} (the usual change of
phases of half of the orbitals of $H_{3b}$ to have the same sign of the
hoppings for all directions changes the sign of $t_1$). After this mapping, $%
c_{i\uparrow }^{\dagger }$ has the character of an effective electron
creation operator (mainly of O character) over a vacuum state where all
sites carry a Zhang-Rice singlet, and $t_1\sim 0.3-0.4$eV $>0$ results.

\bibitem{esk2}  The effect of the correction term $t_1/U$ has been studied
recently using finite size diagonalization by H. Eskes and R. Eder (preprint
cond-mat / 9609233), with results which agree with ours.        

\bibitem{leu} P.W. Leung and R.J. Gooding, Phys. Rev. B {\bf 52}, R15711 (1995).
\end{references}
\end{document}